\documentclass[twocolumn,english,prl,superscriptaddress]{revtex4}
\usepackage[T1]{fontenc}
\usepackage[latin9]{inputenc}
\usepackage{amssymb}
\usepackage{graphicx}

\makeatletter
\@ifundefined{textcolor}{}
{%
 \definecolor{BLACK}{gray}{0}
 \definecolor{WHITE}{gray}{1}
 \definecolor{RED}{rgb}{1,0,0}
 \definecolor{GREEN}{rgb}{0,1,0}
 \definecolor{BLUE}{rgb}{0,0,1}
 \definecolor{CYAN}{cmyk}{1,0,0,0}
 \definecolor{MAGENTA}{cmyk}{0,1,0,0}
 \definecolor{YELLOW}{cmyk}{0,0,1,0}
}

\usepackage{hyperref}
\hypersetup{colorlinks=true}
\makeatother

\usepackage{babel}
\begin{document}

\title{Simultaneous Spin-Charge Relaxation in Double Quantum Dots}

\author{V. Srinivasa}

\email{vsriniv@umd.edu}

\affiliation{Joint Quantum Institute, University of Maryland, College Park, MD
20742 and National Institute of Standards and Technology, Gaithersburg,
MD 20899}

\author{K. C. Nowack}

\affiliation{Kavli Insitute of Nanoscience, TU Delft, Lorentzweg 1, 2628CJ Delft,
the Netherlands}

\author{M. Shafiei }

\affiliation{Kavli Insitute of Nanoscience, TU Delft, Lorentzweg 1, 2628CJ Delft,
the Netherlands}

\author{L. M. K. Vandersypen}

\affiliation{Kavli Insitute of Nanoscience, TU Delft, Lorentzweg 1, 2628CJ Delft,
the Netherlands}

\author{J. M. Taylor}

\affiliation{Joint Quantum Institute, University of Maryland, College Park, MD
20742 and National Institute of Standards and Technology, Gaithersburg,
MD 20899}
\begin{abstract}
We investigate phonon-induced spin and charge relaxation mediated
by spin-orbit and hyperfine interactions for a single electron confined
within a double quantum dot. A simple toy model incorporating both
direct decay to the ground state of the double dot and indirect decay
via an intermediate excited state yields an electron spin relaxation
rate that varies non-monotonically with the detuning between the dots.
We confirm this model with experiments performed on a GaAs double
dot, demonstrating that the relaxation rate exhibits the expected
detuning dependence and can be electrically tuned over several orders
of magnitude. Our analysis suggests that spin-orbit mediated relaxation
via phonons serves as the dominant mechanism through which the double-dot
electron spin-flip rate varies with detuning. 
\end{abstract}
\maketitle
Controlling individual spins is central to spin-based quantum information
processing \cite{Loss1998,Hanson2007RMP,Hanson2008} and also enables
precision metrology \cite{Taylor2008,Dolde2011}. While rapid control
can be achieved by coupling the spins of electrons in semiconductor
quantum dots \cite{Loss1998,Hanson2007RMP} to electric fields via
the electronic charge state \cite{Hanson2008,Kato2003,Rashba2003,Taylor2005,Tokura2006,Nowack2007,Laird2007,Pioro-Ladriere2008,Schreiber2011,Shafiei2013},
spin-charge coupling also leads to relaxation of the spins through
fluctuations in their electrostatic environment. Phonons serve as
an inherent source of fluctuating electric fields in quantum dots
\cite{Hanson2007RMP} and give rise to both charge and spin relaxation
through the electron-phonon interaction. In GaAs quantum dots, the
direct coupling of spin to the strain field produced by phonons is
expected to be inefficient \cite{Khaetskii2000,Khaetskii2001}. The
dominant mechanisms of phonon-induced spin relaxation are therefore
indirect and involve spin-charge coupling due to primarily spin-orbit
\cite{Khaetskii2000,Khaetskii2001,Golovach2004PRL,Bulaev2005,Stano2005,Stano2006PRL,Amasha2008}
and hyperfine \cite{Erlingsson2001,Erlingsson2002,Merkulov2002,Johnson2005,Koppens2005,Taylor2007}
interactions. Confining an electron within a double quantum dot (DQD)
provides a high degree of control over the charge state \cite{vanderWiel2002,Hayashi2003,Petta2004,Gorman2005},
so that relaxation rates can be varied over multiple orders of magnitude
by adjusting the energy level detuning between the dots \cite{Fujisawa1998,Johnson2005,Raith2012,Wang2006,Wang2011}. 

Here, we investigate the interplay of spin and charge relaxation via
phonons for a single electron confined to a DQD in the presence of
spin-orbit and hyperfine interactions. We present a simple model together
with measurements of the electron spin relaxation rate in a GaAs DQD,
both of which yield a non-monotonic dependence on the detuning between
the dots (see Fig. \ref{fig:compexpttheory}). The experiments provide
confirmation of the model and demonstrate the existence of spin \textquotedblleft{}hot
spot\textquotedblright{} features \cite{Fabian1998,Bulaev2005,Stano2005,Stano2006PRL,Raith2011}
at nonzero values of detuning, where relaxation is enhanced by several
orders of magnitude. The opposite behavior is observed at zero detuning,
where the spin-flip rate exhibits a local minimum. Theoretically,
spin hot spots are predicted to occur due to the complete mixing of
spin and orbital states at avoided energy crossings associated with
spin-orbit coupling \cite{Fabian1998,Bulaev2005,Stano2005}. From
a practical standpoint, adjusting the detuning to these points represents
a potential method for rapid all-electrical spin initialization. 

\begin{figure}[b]
\includegraphics[width=2.75in]{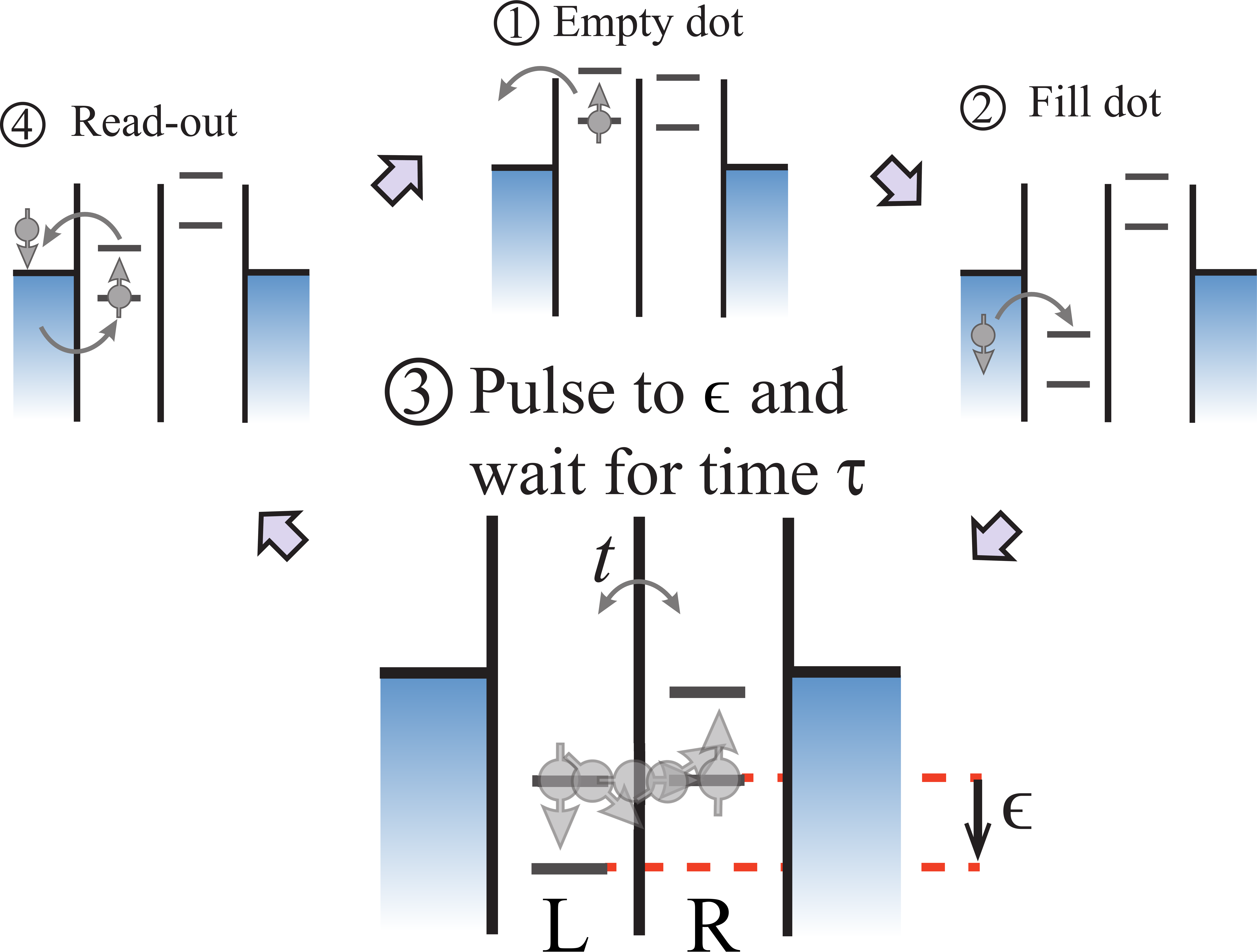}

\caption{\label{fig:sedqdmeas} Electrochemical potential diagrams for a double
quantum dot (DQD), illustrating the measurement cycle used to obtain
the experimental spin relaxation rate (see main text). Varying the
detuning $\epsilon$ between the left (L) and right (R) dots while
keeping the tunnel coupling $t$ fixed (stage 3) tunes the relative
energies of the charge states. Tunneling of the electron between the
dots is accompanied by spin rotation.}
\end{figure}

\begin{figure}
\includegraphics[bb=10bp 300bp 500bp 820bp,clip,width=3.375in]{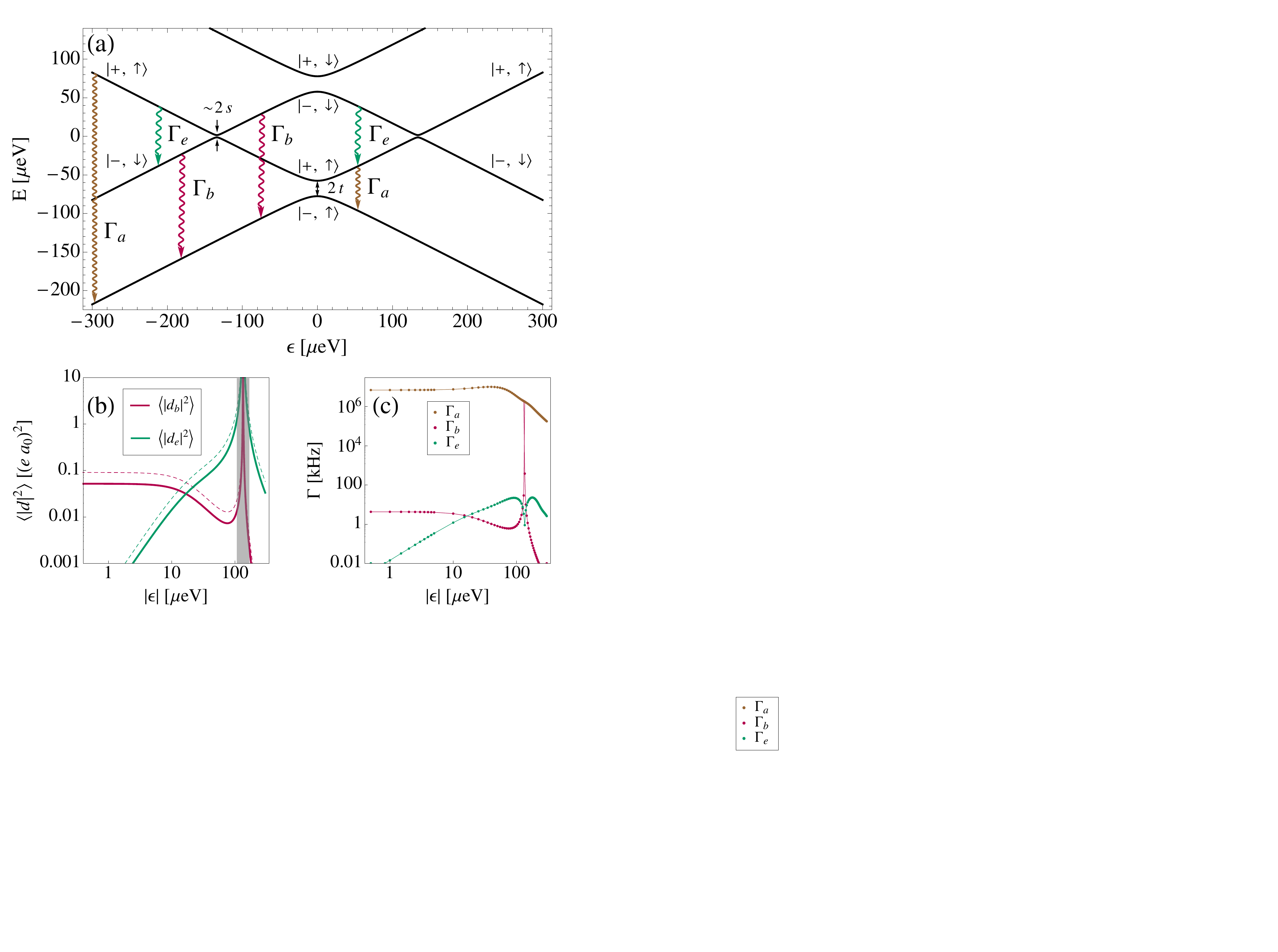}

\caption{\label{fig:spinrelax} (a) Spectrum of $H_{d}$ {[}Eq. (\ref{eq:hd}){]}
as a function of detuning $\epsilon$ for the case $t\ll\Omega_{z},$
in the presence of spin-orbit coupling ($\mathbf{K}_{\mathrm{so}}\neq\mathbf{0},$
$\mathbf{K}_{\mathrm{nuc}}=\mathbf{0}$). Correspondence with the
eigenstates of $H_{0}$ {[}Eqs. (\ref{eq:h0}) and (\ref{eq:pmstates}){]}
is indicated for relevant regions of the spectrum. Avoided crossings
due to spin-orbit coupling occur at $\epsilon\approx\pm\Omega_{z}$.
The spectrum shown corresponds to $t=10\ \mu\mbox{eV}$ \cite{SupplMat},
$B=6.5\ \mbox{T}$ \cite{Nowack2011,SupplMat}, the dot size $\sigma=15\ \mbox{nm}$
and interdot separation $a=110\ \mbox{nm},$ the GaAs effective mass
$m^{*}=0.067m_{e}$ (where $m_{e}$ is the free-electron mass) and
g-factor $g=0.36$, the Rashba and linear Dresselhaus spin-orbit strengths
$\alpha_{0}=3.3\times10^{-12}\ \mbox{eV\ensuremath{\cdot}m}$ and
$\beta_{0}=4.5\times10^{-12}\ \mbox{eV\ensuremath{\cdot}m}$, respectively,
and $\theta=\pi/8.$ The values of $\alpha_{0},$ $\beta_{0},$ and
$\theta$ are used to determine $\mathbf{K}_{\mathrm{so}}=\left(r,s,q\right).$
(b) Dipole-dependent factors $\langle|d_{b}|^{2}\rangle$ and $\langle|d_{e}|^{2}\rangle$
{[}Eqs. (\ref{eq:dplussqavg}) and (\ref{eq:desqavg}){]} used to
qualitatively model the relaxation rates $\Gamma_{b}$ and $\Gamma_{e}$
in (a), as a function of detuning for $B_{\mathrm{nuc}}=0$ (solid
lines) and $B_{\mathrm{nuc}}=3\ \mathrm{mT}$ (dotted lines), with
$\alpha_{0}=3.3\times10^{-14}\ \mbox{eV\ensuremath{\cdot}m}$ and
$\beta_{0}=4.5\times10^{-14}\ \mbox{eV\ensuremath{\cdot}m}$. All
other parameters are identical to those used in (a). Units for the
dipole are given in terms of the Bohr radius $a_{0}.$ The dipole
model is not valid in the shaded region. (c) Relaxation rates $\Gamma_{a},$
$\Gamma_{b},$ and $\Gamma_{e}$ as a function of detuning. The rates
are calculated using $\rho_{0}=5.3\times10^{3}\ \mbox{kg/m}^{3},$
$c_{l}=5.3\times10^{3}\ \mathrm{m}/\mathrm{s},$ $c_{t}=2.5\times10^{3}\ \mathrm{m}/\mathrm{s},$
$\beta_{l}=7.0\ \mbox{eV},$ and $\Xi=1.4\times10^{9}\ \mbox{eV/m}$
\cite{Stano2006PRL}, together with the parameter values used in (b).
Lines are guides to the eye. }
\end{figure}

\begin{figure}
\includegraphics[bb=40bp 30bp 450bp 330bp,width=3in]{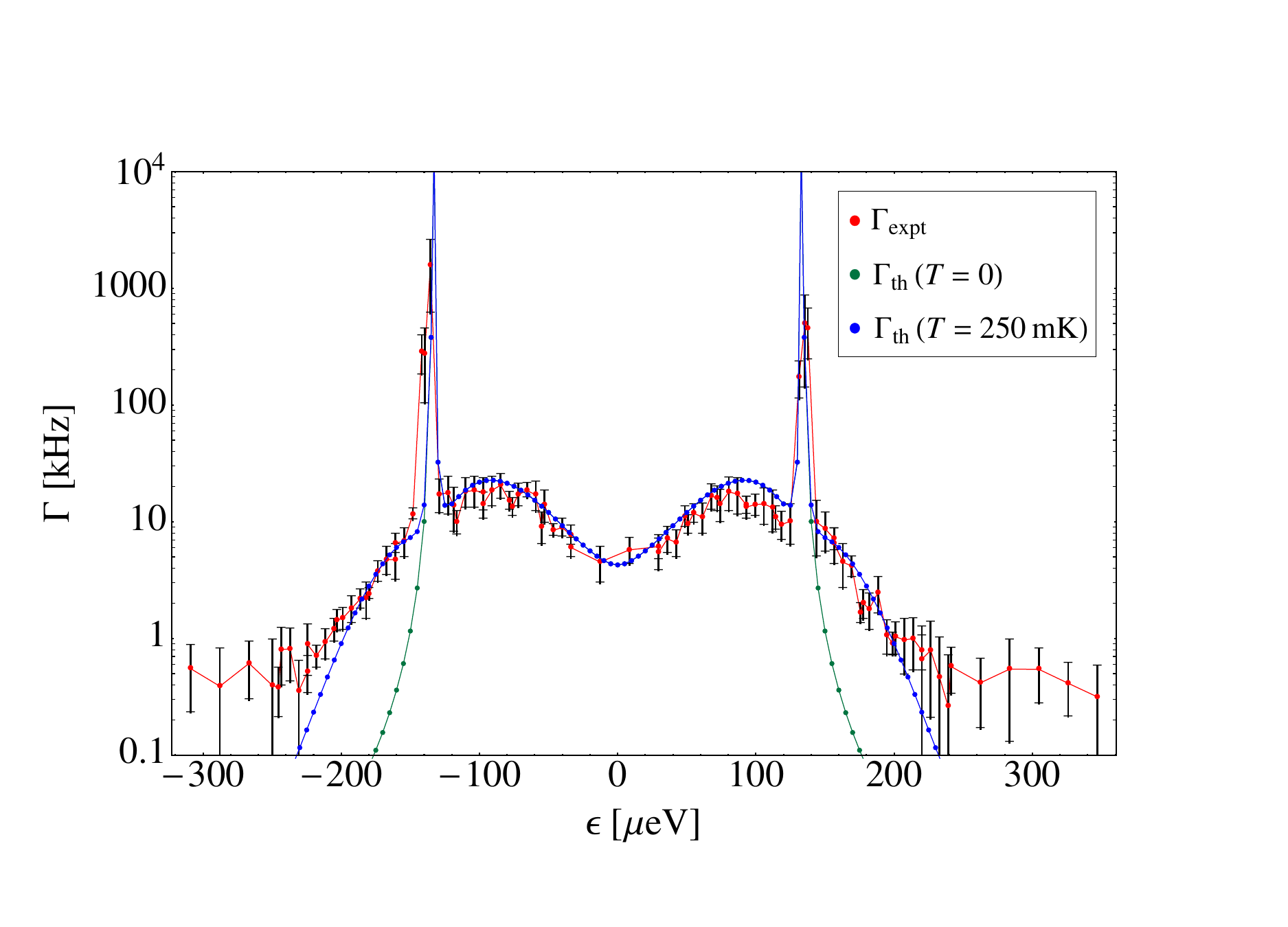}

\caption{\label{fig:compexpttheory} Experimental detuning-dependent single-spin
relaxation rate $\left(\Gamma_{\mathrm{expt}}\right)$ and comparison
with the toy model described in the present work $\left(\Gamma_{\mathrm{th}}\right)$
for both zero and finite temperature. Error bars depict 90\% confidence
intervals for the data. The parameter values used to calculate $\Gamma_{\mathrm{th}}$
are the same as those used in (b) and (c) of Fig. \ref{fig:spinrelax}.}
\end{figure}

We describe a single electron confined within a DQD (Fig. \ref{fig:sedqdmeas})
using a toy model that includes only the lowest-energy orbital level
of each dot. This two-level approximation \cite{vanderWiel2002} enables
the charge degrees of freedom to be represented by the Pauli matrices
$\sigma_{x},\sigma_{y},\sigma_{z}$ in the basis $\left\{ \left|L\right\rangle ,\left|R\right\rangle \right\} $,
where $\left|L\right\rangle $ ($\left|R\right\rangle $) denotes
an electron in the left (right) quantum dot and $\sigma_{z}\equiv\left|L\right\rangle \left\langle L\right|-\left|R\right\rangle \left\langle R\right|$.
We can then express the Hamiltonian of the system as

\begin{eqnarray}
H_{d} & = & H_{\mathrm{0}}+H_{\mathrm{so}}+H_{\mathrm{nuc}},\label{eq:hd}\\
H_{\mathrm{0}} & = & \frac{\epsilon}{2}\sigma_{z}-t\sigma_{x}-\Omega_{z}S_{z},\label{eq:h0}\\
H_{\mathrm{so}} & = & \mathbf{K}_{\mathrm{so}}\cdot\mathbf{S}\ \sigma_{y},\label{eq:hso}\\
H_{\mathrm{nuc}} & = & \mathbf{K}_{\mathrm{nuc}}\cdot\mathbf{S}\ \sigma_{z}.\label{eq:hnuc}
\end{eqnarray}
The first two terms in $H_{0}$ {[}Eq. (\ref{eq:h0}){]} specify the
purely orbital part $H_{\mathrm{orb}}\equiv\left(\epsilon/2\right)\sigma_{z}-t\sigma_{x}$
of the electronic Hamiltonian in terms of the energy level detuning
$\epsilon$ and the tunnel coupling $t$ between the two dots (Fig.
\ref{fig:sedqdmeas}). Diagonalization of $H_{\mathrm{orb}}$ yields
the eigenstates
\begin{eqnarray}
\left|+\right\rangle  & \equiv & \cos\frac{\phi}{2}\left|L\right\rangle -\sin\frac{\phi}{2}\left|R\right\rangle ,\nonumber \\
\left|-\right\rangle  & \equiv & \sin\frac{\phi}{2}\left|L\right\rangle +\cos\frac{\phi}{2}\left|R\right\rangle ,\label{eq:pmstates}
\end{eqnarray}
where $\phi$ varies with $\epsilon$ and $t$ according to $\tan\phi=2t/\epsilon.$
The corresponding eigenvalues are separated in energy by a gap $\Delta=E_{+}-E_{-}=\sqrt{\epsilon^{2}+4t^{2}}$
{[}see also Fig. \ref{fig:spinrelax}(a){]}. 

Spin dependence is introduced into the Hamiltonian via the last term
in $H_{0},$ together with $H_{\mathrm{so}}$ and $H_{\mathrm{nuc}}$
{[}see Eqs. (\ref{eq:h0})-(\ref{eq:hnuc}){]}, where the vector of
electron spin operators is denoted by $\mathbf{S}=\left(S_{x},\ S_{y},\ S_{z}\right)$.
The term $H_{\mathrm{so}}$ describes spin-orbit coupling which is
linear in the electron momentum $\mathbf{p}$. The general form given
in Eq. (\ref{eq:hso}) takes into account both the Rashba \cite{Rashba1960,Bychkov1984}
and the linear Dresselhaus \cite{Dresselhaus1955} forms of spin-orbit
interaction, with strengths and orientations that are specified by
the vector $\mathbf{K}_{\mathrm{so}}\equiv\left(r,s,q\right)$. Note
that $H_{\mathrm{so}}$ acts as $\sigma_{y}$ within the orbital subspace,
which follows from parity selection rules for the matrix elements
of $\mathbf{p}$ in the basis $\left\{ \left|L\right\rangle ,\left|R\right\rangle \right\} $.
Thus, $H_{\mathrm{so}}$ describes tunneling between the dots accompanied
by a spin flip (Fig. \ref{fig:sedqdmeas}) \cite{Schreiber2011}. 

The remaining spin-dependent terms in $H_{d}$ represent forms of
the Zeeman interaction that are distinguished by their action within
the orbital subspace. The final term in $H_{0}$ represents the coupling
of the electron spin to a magnetic field of strength $B=\Omega_{z}/\left|g\right|\mu_{B}$
that is uniform over the two dots, where $g$ is the electron g-factor
and $\mu_{B}$ is the Bohr magneton. The vector $\mathbf{K}_{\mathrm{nuc}}\equiv\left(u,v,w\right)$
in $H_{\mathrm{nuc}}$ specifies the strength and orientation of a
magnetic field gradient across the two dots. $H_{\mathrm{nuc}}$ acts
as $ $$\sigma_{z}$ within the orbital subspace. For GaAs quantum
dots, $H_{\mathrm{nuc}}$ can be used to model the hyperfine interaction
between the electron spin and the ensemble of nuclear spins with which
the electron wavefunction overlaps. The associated intrinsic magnetic
field gradient is assumed to originate from an effective nuclear field
$\mathbf{B}_{\mathrm{nuc}}$ with a random, spatially-varying orientation
described by a Gaussian distribution and magnitude $B_{\mathrm{nuc}}$
given by the root-mean-square (rms) value \cite{Merkulov2002,Erlingsson2002,Johnson2005,Taylor2007,Noteunifnuc}.

Figure \ref{fig:sedqdmeas} illustrates the scheme used for the measurement
of the spin relaxation rate. The experimental setup is described in
\cite{SupplMat}. In the first step of the measurement cycle, a single
electron spin is initialized by emptying the DQD and then letting
a single electron tunnel into the left dot far from the degeneracy
of $\left|L\right\rangle $ and $\left|R\right\rangle $. The electron
spin is randomly up or down. Next, a voltage pulse adjusts the electrochemical
potential of the right dot to tune the detuning closer to the degeneracy
to a value $\epsilon$ for a wait time $\tau$. After the wait time,
the electrochemical potential is pulsed back and the spin of the electron
is read out using energy-selective spin-to-charge conversion \cite{Elzerman2004}.
This cycle is repeated for a given $\epsilon$ and $\tau$ to obtain
an average spin-down probability at the end of the cycle. For each
series of measurements as a function of $\tau$ at a fixed $\epsilon,$
the amount of detected spin-down is fitted with an exponential decay,
from which the spin-relaxation rate at each $\epsilon$ is obtained
as shown in Fig. \ref{fig:compexpttheory}. 

The variation of the measured spin relaxation rate with detuning can
be understood in terms of the spectrum for the one-electron double
dot. Figure \ref{fig:spinrelax}(a) shows an example spectrum for
$H_{d}$ {[}Eq. (\ref{eq:hd}){]} as a function of detuning. In Fig.
\ref{fig:spinrelax} and throughout the present work, we consider
the limit $t\ll\Omega_{z}$ which corresponds to the measurements
described above (see \cite{SupplMat}). The notation $\uparrow,\downarrow$
used to label the states in the figure refers to the components of
spin along the quantization axis defined by the external magnetic
field. In accordance with the experiment \cite{SupplMat,Nowack2011},
we choose this field to be in the plane of the quantum dots and parallel
to the double-dot axis. The in-plane crystal lattice orientation characterizing
the spin-orbit interaction {[}Eq. (\ref{eq:hso}){]} is parametrized
by an angle $\theta$ relative to this axis. Of particular consequence
for the present work is the fact that $H_{\mathrm{so}}$ gives rise
to avoided crossings in the spectrum at $\epsilon\approx\pm\Omega_{z},$
where maximal coupling of the states $\left|+,\uparrow\right\rangle $
and $\left|-,\downarrow\right\rangle $ occurs and leads to the complete
mixing of orbital and spin degrees of freedom. These finite values
of $\epsilon$ correspond to spin ``hot spots'' \cite{Fabian1998,Bulaev2005,Stano2005,Stano2006PRL,Raith2011}
and are associated with enhanced spin relaxation rates, as is shown
below. 

Including coupling to phonons in the description of the single-electron
double-dot system leads to the Hamiltonian $H=H_{d}+H_{\mathrm{ep}},$
where 
\begin{eqnarray}
H_{\mathrm{ep}} & = & \sum_{\nu,\mathbf{k}}\sqrt{\frac{\hbar}{2\rho_{0}V_{0}c_{\nu}k}}\left(k\beta_{l}\delta_{\nu,l}-i\Xi\right)\nonumber \\
 &  & \ \ \ \ \ \times\left(a_{\nu,\mathbf{k}}+\ a_{\nu,-\mathbf{k}}^{\dagger}\right)e^{i\mathbf{k}\cdot\mathbf{r}}\label{eq:hep}
\end{eqnarray}
is the electron-phonon interaction \cite{Mahan2000}, expressed in
terms of the mass density $\rho_{0}$, the volume $V_{0}$, the phonon
speeds $c_{\nu}$, the deformation potential $\beta_{l},$ and the
piezoelectric constant $\Xi$. The operator $a_{\nu,\mathbf{k}}^{\dagger}$
($a_{\nu,\mathbf{k}}$) creates (annihilates) a phonon with wavevector
$\mathbf{k}$ and polarization $\nu$ {[}the sum over $\nu$ is taken
over one longitudinal $\left(l\right)$ mode and two transverse $\left(t\right)$
modes{]}, and $\delta_{\nu,l}$ is the Kronecker delta function. Fermi's
golden rule for the rate $\Gamma$ of phonon-induced relaxation of
the electron from state $\left|i\right\rangle $ to state $\left|f\right\rangle $$ $
of the double dot gives $\Gamma\sim\left|\left\langle f\right|H_{\mathrm{ep}}\left|i\right\rangle \right|^{2}\rho\left(\Delta_{d}\right)$,
where $\rho\left(\Delta_{d}\right)$ is the phonon density of states
evaluated at the gap $\Delta_{d}$ between levels $i$ and $f$ that
determines the energy of the emitted phonon. 

We first consider a qualitative model for $\Gamma$, where we estimate
the transition matrix element $\left\langle f\right|e^{i\mathbf{k}\cdot\mathbf{r}}\left|i\right\rangle $
(see \cite{SupplMat}) by writing $e^{i\mathbf{k}\cdot\mathbf{r}}\approx1+i\mathbf{k}\cdot\mathbf{r}$
and determining the corresponding matrix element of the dipole operator
$\mathbf{d}=-e\mathbf{r}$ (here, $e$ denotes the magnitude of the
electron charge). To evaluate dipole matrix elements, we define Gaussian
wavefunctions $\psi_{L\left(R\right)}\left(\mathbf{r}\right)\equiv\left\langle \mathbf{r}\right|\left.L\left(R\right)\right\rangle $
which are shifted along the dot axis by $\pm a/2$ for the left-localized
and right-localized orbital states. While $\psi_{L}$ and $\psi_{R}$
are not orthogonal, their overlap is small. We neglect corrections
due to this overlap in our calculations. Using these wavefunctions,
we find $\mathbf{d}=D\hat{z}$ with $D=\left(ea/2\right)\sigma_{z}$.
The qualitative form of the relaxation rate can then be approximated
by $\Gamma\sim\left|d\right|^{2}F\left(\Delta_{d}\right),$ where
$d$ denotes the first-order term of $\left\langle f\right|D\left|i\right\rangle $
and $F\left(\Delta_{d}\right)$ represents the dependence of the rate
on the gap energy $\Delta_{d}$ (see \cite{SupplMat} for more details). 

To identify the states of the double dot between which phonon-induced
relaxation occurs, we treat $V\equiv H_{\mathrm{so}}+H_{\mathrm{nuc}}$
{[}Eqs. (\ref{eq:hso}) and (\ref{eq:hnuc}){]} as a perturbation
with respect to $H_{0}$ {[}Eq. (\ref{eq:h0}){]} and use nondegenerate
perturbation theory (which is valid away from $\epsilon\approx\pm\Omega_{z}$)
to find the first-order corrections to the energies and eigenstates
of $H_{0}$. We denote the corrected states by $\left\{ \tilde{\left|-,\uparrow\right\rangle },\tilde{\left|-,\downarrow\right\rangle },\tilde{\left|+,\uparrow\right\rangle },\tilde{\left|+,\downarrow\right\rangle }\right\} $
and consider relaxation of the electron spin from the excited state
$\tilde{\left|-,\downarrow\right\rangle }$ to the ground state $\tilde{\left|-,\uparrow\right\rangle }$
of the DQD {[}see Fig. \ref{fig:spinrelax}(a){]}, which can occur
directly as well as indirectly via the excited state $\tilde{\left|+,\uparrow\right\rangle }.$
Away from the avoided crossing points, we note that $\tilde{\left|+,\uparrow\right\rangle }\approx\left|+,\uparrow\right\rangle $
and $\tilde{\left|-,\uparrow\right\rangle }\approx\left|-,\uparrow\right\rangle $.
The state $\tilde{\left|+,\uparrow\right\rangle }$ therefore relaxes
rapidly to $\tilde{\left|-,\uparrow\right\rangle }$, as effectively
only orbital decay is involved and no spin flip is required in this
second step \cite{Fujisawa2001}. In the following, we assume that
this charge relaxation is instantaneous compared to the spin relaxation
and use the dipole matrix element for $\tilde{\left|-,\downarrow\right\rangle }\rightarrow\tilde{\left|+,\uparrow\right\rangle }$
to describe the full indirect transition. 

We approximate the relaxation rates $\Gamma_{b}$ and $\Gamma_{e}$
{[}Fig. \ref{fig:spinrelax}(a){]} in the presence of both $H_{\mathrm{so}}$
and $H_{\mathrm{nuc}}$ by calculating the first-order terms $d_{b}$
and $d_{e}$ of the dipole matrix elements $\tilde{\left\langle -,\uparrow\right|}D\tilde{\left|-,\downarrow\right\rangle }$
and $\tilde{\left\langle +,\uparrow\right|}D\tilde{\left|-,\downarrow\right\rangle }$,
respectively. These terms are functions of the spin-flipping components
$r,\ s,\ u,$ and $v$ in Eqs. (\ref{eq:hso}) and (\ref{eq:hnuc}).
Averaging over the nuclear field distribution \cite{Merkulov2002,Erlingsson2002,Johnson2005,Taylor2007}
\begin{eqnarray}
P\left(\mathbf{K}_{\mathrm{nuc}}\right) & = & \frac{1}{\left(2\pi b_{\mathrm{nuc}}^{2}\right)^{3/2}}\exp\left(-\frac{\left|\mathbf{K}_{\mathrm{nuc}}\right|^{2}}{2b_{\mathrm{nuc}}^{2}}\right),\label{eq:pnuc}
\end{eqnarray}
where $b_{\mathrm{nuc}}\equiv\left|g\right|\mu_{B}B_{\mathrm{nuc}}=\sqrt{\langle\left|\mathbf{K}_{\mathrm{nuc}}\right|^{2}\rangle/3},$
gives $\langle u\rangle=\langle v\rangle=0$ and $\langle u^{2}\rangle=\langle v^{2}\rangle=b_{\mathrm{nuc}}^{2}.$
We thus find 
\begin{eqnarray}
\langle|d_{b}|^{2}\rangle & = & \left[\frac{ea}{2}\left(\frac{2t}{\Delta}\right)\frac{\Omega_{z}}{(\Delta-\Omega_{z})(\Delta+\Omega_{z})}\right]^{2}\chi,\label{eq:dplussqavg}\\
\langle|d_{e}|^{2}\rangle & = & \left[\frac{ea}{2}\left(\frac{\epsilon}{\Delta}\right)\frac{1}{\Delta-\Omega_{z}}\right]^{2}\chi,\label{eq:desqavg}\\
\chi & \equiv & \left[r^{2}+s^{2}+\left(\frac{2t}{\Omega_{z}}\right)^{2}\left(2b_{\mathrm{nuc}}^{2}\right)\right].\nonumber 
\end{eqnarray}
These expressions are plotted in Fig. \ref{fig:spinrelax}(b). Note
that both Eqs. (\ref{eq:dplussqavg}) and (\ref{eq:desqavg}) are
undefined at the avoided crossing points in Fig. \ref{fig:spinrelax}(a),
where $\Delta=\Omega_{z}$. Thus, the curves shown in Fig. \ref{fig:spinrelax}(b)
are valid only away from these points (i.e., where nondegenerate perturbation
theory is a reasonable approximation). Both $\langle|d_{b}|^{2}\rangle$
and $\langle|d_{e}|^{2}\rangle$ are only slightly modified by the
coupling of the electron spin to an effective nuclear field of rms
strength $B_{\mathrm{nuc}}=3\ \mathrm{mT}$ \cite{Johnson2005}, as
expected from Eqs. (\ref{eq:dplussqavg}) and (\ref{eq:desqavg})
in which the nuclear field term is scaled with respect to the spin-orbit
terms by a factor $\left(2t/\Omega_{z}\right)^{2}\ll1$ \cite{Schreiber2011}.
Saturation of $\langle|d_{b}|^{2}\rangle$ occurs at zero detuning
for both the $B_{\mathrm{nuc}}=0$ and the $B_{\mathrm{nuc}}=3\ \mathrm{mT}$
cases. On the other hand, $\langle|d_{e}|^{2}\rangle$ vanishes at
$\epsilon=0$ regardless of the value of $B_{\mathrm{nuc}}$. As the
experimental relaxation rate contains a local minimum at zero detuning
(see Fig. \ref{fig:compexpttheory}), the present analysis suggests
that the direct transition $\tilde{\left|-,\downarrow\right\rangle }\rightarrow\tilde{\left|-,\uparrow\right\rangle }$
alone does not account for the observed spin relaxation and that indirect
relaxation via the excited state $\tilde{\left|+,\uparrow\right\rangle }$
potentially plays a significant role in the spin-flip rate. The relative
contributions of the direct and indirect transitions to the overall
rate are explored in \cite{SupplMat}.

To compare our theoretical predictions more directly with the experimental
results, we carry out the full calculation of the relaxation rates
for both direct and indirect transitions to the ground state by applying
Fermi's golden rule to relaxation induced by $H_{\mathrm{ep}}$ {[}Eq.
(\ref{eq:hep}){]}. Details are given in the Supplemental Material
\cite{SupplMat}. We set $\mathbf{K}_{\mathrm{nuc}}=0$ for simplicity,
as the preceding analysis based on dipole matrix elements suggests
that the hyperfine term $H_{\mathrm{nuc}}$ represents only a small
correction to the decay rate {[}see Eqs. (\ref{eq:dplussqavg}), (\ref{eq:desqavg})
and Fig. \ref{fig:spinrelax}(b){]}. Application of a Schrieffer-Wolff
transformation \cite{Schrieffer1966} enables diagonalization of the
full double-dot Hamiltonian $H_{d}$ for all $\epsilon,$ including
the avoided crossing points $\epsilon\approx\pm\Omega_{z}$, and the
eigenstates of $H_{d}$ are used to calculate the relaxation rates
via Eqs. (S1)-(S3) \cite{SupplMat}.

Relevant portions of the curves for the decay rates $\Gamma_{21},$
$\Gamma_{31},$ and $\Gamma_{32}$ (where we number the levels according
to their energy eigenvalues and use $\Gamma_{if}$ to denote the rate
of relaxation from level $i$ to level $f$) are plotted together
in Fig. $ $\ref{fig:spinrelax}(c). The rate $\Gamma_{a}$ is associated
with mainly charge relaxation and is given by $\Gamma_{21}$ ($\Gamma_{31}$)
for $\left|\epsilon\right|\lesssim\Omega_{z}$ ($\left|\epsilon\right|\gtrsim\Omega_{z}$),
while $\Gamma_{b}$ is associated with mainly spin relaxation and
is given by $\Gamma_{31}$ ($\Gamma_{21}$) for $\left|\epsilon\right|\lesssim\Omega_{z}$
($\left|\epsilon\right|\gtrsim\Omega_{z}$). The rate $\Gamma_{e}$
corresponds to a combination of spin and charge relaxation and is
given by $\Gamma_{32}$ for all $\left|\epsilon\right|$. Note that
$\Gamma_{a}\gg\Gamma_{e},$ which is consistent with our prior assumption
that the effective rate for indirect relaxation to the ground state
is determined by $\Gamma_{e}$.

For $\left|\epsilon\right|\lesssim\Omega_{z},$ indirect spin relaxation
occurs by a transition to the lower-energy intermediate state via
phonon emission {[}Fig. \ref{fig:spinrelax}(a){]}. On the other hand,
the indirect transition for $\left|\epsilon\right|\gtrsim\Omega_{z}$
requires phonon absorption in order to excite the electron to the
higher-energy intermediate state. Using the Einstein coefficients
and the Bose-Einstein distribution $\left\langle n\right\rangle =1/\left[\exp\left(\Delta_{d}/k_{B}T\right)-1\right]$
(where $k_{B}$ is the Boltzmann constant and $T$ is the temperature)
to express the rates of spontaneous emission, stimulated emission,
and absorption associated with the lowest three double-dot levels
in Fig. \ref{fig:spinrelax}(a) in terms of $\Gamma_{a},$ $\Gamma_{b},$
and $\Gamma_{e}$ \cite{Fujisawa1998}, we take the full theoretical
$ $detuning-dependent spin relaxation rate $\Gamma_{\mathrm{th}}$
to be given by $\Gamma_{b}+\Gamma_{e}$ for $\left|\epsilon\right|\lesssim\Omega_{z}$
and by $\Gamma_{b}+\Gamma_{e}\left\langle n\right\rangle /\left(\left\langle n\right\rangle +1\right)$
for $\left|\epsilon\right|\gtrsim\Omega_{z}.$ This rate is plotted
together with the zero-temperature limit of the model and the measured
rate $\Gamma_{\mathrm{expt}}$ in Fig. \ref{fig:compexpttheory} for
$T=250\ \mathrm{mK}$ \cite{SupplMat,Nowack2011}. At the avoided
crossings associated with spin-orbit coupling ($\epsilon\approx\pm\Omega_{z}$),
we find peaks in $\Gamma_{\mathrm{th}}$ that resemble the spin hot
spot peaks observed experimentally. The zero-detuning minimum found
in the measurements appears in both the zero-and the finite-temperature
models. In addition, close qualitative agreement between the finite-temperature
model and experiment is observed for a wide range of detuning values.
While limitations of our theoretical description arise from the two-level
approximation we use for the orbital states, we have nevertheless
shown that several characteristic features present in the measured
detuning dependence of the double-dot spin relaxation rate can be
understood within this simple model. 

The results of the present work therefore suggest that, in accordance
with the case of single lateral GaAs quantum dots \cite{Amasha2008},
the observed variation of the spin relaxation rate with detuning in
double dots is dominated by spin-orbit mediated electron-phonon coupling.
The spin-orbit interaction may thus provide the key to rapid all-electrical
initialization of single spins.

\emph{Note added}. During the preparation of this manuscript, we became
aware of a recent experimental observation of a spin hot spot in a
Si quantum dot \cite{Yang2013}. 

We thank N. M. Zimmerman, M. D. Stiles, and P. Stano for helpful comments.
Research was supported by DARPA MTO, the Office of the Director of
National Intelligence, Intelligence Advanced Research Projects Activity
(IARPA), through the Army Research Office (Grant W911NF-12-1-0354),
SOLID (EU), and an ERC Starting Grant.

\bibliographystyle{apsrev4-1}
\bibliography{ECSRDQD}

\begin{thebibliography}{48}
\bibitem [{Not()}]{Noteunifnuc}%
  \BibitemOpen
  \href@noop {} {}\bibinfo {howpublished} {Note that $\Omega_{z}$ implicitly
  includes the homogeneous part of the nuclear field.}\BibitemShut {Stop}%
\end{thebibliography}

\end{document}


\title{Supplemental Material: Simultaneous Spin-Charge Relaxation in Double
Quantum Dots}

\author{V. Srinivasa}

\affiliation{Joint Quantum Institute, University of Maryland, College Park, MD
20742 and National Institute of Standards and Technology, Gaithersburg,
MD 20899}

\author{K. C. Nowack}

\affiliation{Kavli Insitute of Nanoscience, TU Delft, Lorentzweg 1, 2628CJ Delft,
the Netherlands}

\author{M. Shafiei }

\affiliation{Kavli Insitute of Nanoscience, TU Delft, Lorentzweg 1, 2628CJ Delft,
the Netherlands}

\author{L. M. K. Vandersypen}

\affiliation{Kavli Insitute of Nanoscience, TU Delft, Lorentzweg 1, 2628CJ Delft,
the Netherlands}

\author{J. M. Taylor}

\affiliation{Joint Quantum Institute, University of Maryland, College Park, MD
20742 and National Institute of Standards and Technology, Gaithersburg,
MD 20899}

\maketitle

\section{Spin relaxation rate measurement}

\begin{figure}
\includegraphics[bb=10bp 0bp 168bp 220bp,width=1.8in]{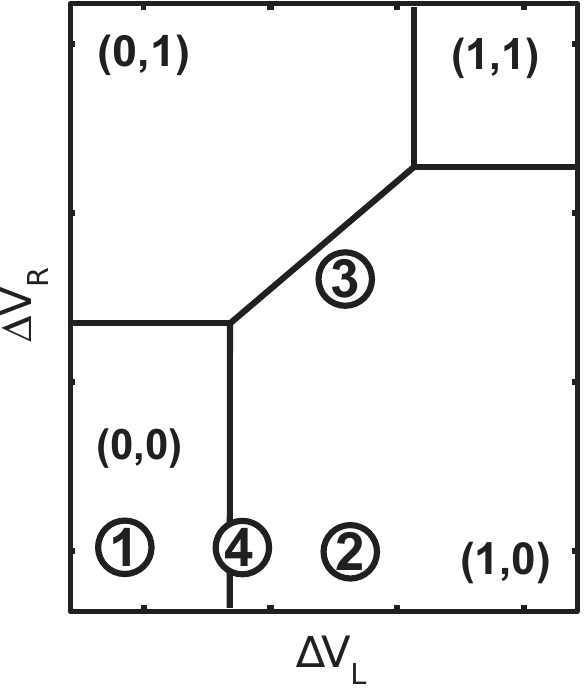}

\caption{\label{chargingdiag}Schematic of the double quantum dot charge stability
diagram associated with the steps in the measurement scheme of Fig.
1 in the main text. $\left(n,m\right)$ represents a state with $n$
electrons in the left dot and $m$ electrons in the right dot. The
charge configurations $\left(1,0\right)$ and $\left(0,1\right)$
correspond to the states $\left|L\right\rangle $ and $\left|R\right\rangle $,
respectively, introduced in the main text. $\Delta V_{L}$ and $\Delta V_{R}$
represent the effective voltages applied to the left and right gates. }
\end{figure}

The measurement illustrated in Fig. 1 of the main text is performed
on a DQD formed by using Ti/Au surface gates to locally deplete a
two-dimensional electron gas 90 nm below the surface of a GaAs/(Al,Ga)As
heterostructure. Quantum point contacts (QPCs) on both sides of the
DQD are used to detect the charge on each dot. We tune the DQD to
contain a single electron (Fig. \ref{chargingdiag}) and operate close
to the degeneracy of $\left|L\right\rangle $ and $\left|R\right\rangle $.
To set the electrochemical potentials in the left and right dot independently,
combinations of voltages on two gates, one closer to the left dot
and the other closer to the right dot, are tuned which compensate
for capacitive cross-coupling. The tunnel coupling is tuned to be
approximately $8\ \mu\n{eV}$. Both the tunnel coupling and the conversion
from applied gate voltage to change in detuning are determined from
microwave spectroscopy \cite{Petta2004}. An in-plane magnetic field
$B_{\n{ext}}=6.5\ \n{T}$ is applied to split the spin-up and spin-down
energy of the electron by the Zeeman energy ($E_{\n{Z}}\approx130\ \mu\n{eV}$).
The electron temperature is typically 250 mK \cite{Nowack2011}.

\section{Comparison of direct and indirect spin relaxation rates}

The relative contributions of the direct and indirect transitions
described in the main text to the overall relaxation rate are characterized
by the ratio $\Gamma_{e}/\Gamma_{b}.$ An estimate of this ratio can
be obtained by combining the dipole-dependent factors $\langle|d_{b}|^{2}\rangle$
and $\langle|d_{e}|^{2}\rangle$ given in Eqs. (8) and (9) of the
main text with the corresponding energy-dependent factors $F\left(\Delta_{b}\right)$
and $F\left(\Delta_{e}\right),$ where $\Delta_{b}\approx\Omega_{z}$
and $\Delta_{e}\approx\Omega_{z}-\Delta$ are the associated energy
gaps. To determine an approximate form for $F\left(\Delta_{d}\right)$,
we consider the full expression obtained from Fermi's golden rule
for the rate of decay induced by $H_{\mathrm{ep}}$ {[}Eq. (6) of
the main text{]}. Defining $M_{k}\equiv e^{i\mathbf{k}\cdot\mathbf{r}},$
we obtain

\begin{eqnarray}
\Gamma & = & \frac{2\pi}{\hbar}\sum_{\nu}\frac{V_{0}}{(2\pi)^{3}}\int k^{2}dk\frac{\hbar}{2\text{\ensuremath{\rho}}_{0}V_{0}c_{\nu}k}\left(k^{2}\beta_{l}{}^{2}\delta_{\nu,l}\right.\nonumber \\
 &  & \ \ \ \ \ \ \ \ \ \left.+\ \Xi{}^{2}\right)I_{\text{ang}}(k)\delta\left(\varepsilon_{\mathrm{ph}}-\Delta_{d}\right),\label{eq:decayrate}
\end{eqnarray}
where we have converted the sum over $\mathbf{k}$ to an integral,
and 
\begin{equation}
I_{\text{ang}}(k)=\int\text{d\ensuremath{\Omega}}\left|\left\langle f\right|M_{k}\left|i\right\rangle \right|{}^{2}\label{eq:intang}
\end{equation}
is a momentum-space angular integral that is determined by the transition
matrix element $\left\langle f\right|M_{k}\left|i\right\rangle $.
In $H_{\mathrm{ep}}$ and Eq. (\ref{eq:decayrate}), a linear dispersion
$\varepsilon_{\mathrm{ph}}=\hbar c_{\nu}k$ for the phonon energies
is assumed. The only factor in $H_{\mathrm{ep}}$ which depends on
electronic degrees of freedom (specifically, on the position operator
$\mathbf{r}$ for the electron) is $M_{k},$ which can be expressed
within the same two-level approximation used for $H_{d}$ in Eqs.
(1)-(4) of the main text. The form of $M_{k}$ in the basis $\left\{ \left|L\right\rangle ,\left|R\right\rangle \right\} $
is given by 
\begin{eqnarray}
M_{k} & = & e^{-\left(k_{y}^{2}+k_{z}^{2}\right)\sigma^{2}/2}\left[\cos\left(\frac{ak_{z}}{2}\right)\boldsymbol{1}\right.\nonumber \\
 &  & \left.-\ i\sin\left(\frac{ak_{z}}{2}\right)\sigma_{z}+e^{-a^{2}/8\sigma^{2}}\sigma_{x}\right],\label{eq:mk}
\end{eqnarray}
where the term involving $\sigma_{x}$ describes phonon-assisted tunneling
between the left and right dots. 

\begin{figure}
\includegraphics[bb=0bp 30bp 500bp 330bp,width=3.375in]{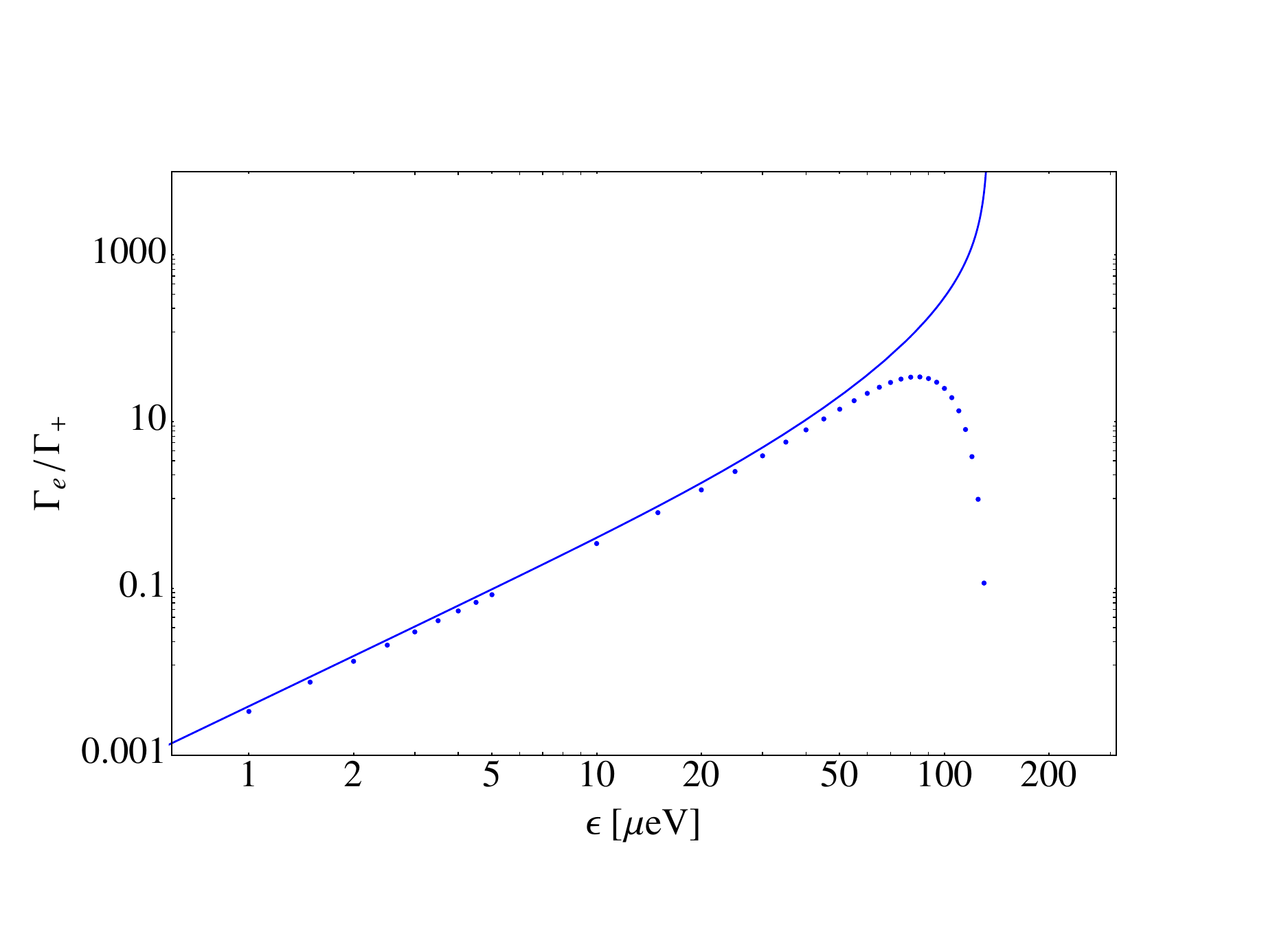}

\caption{\label{decayrateratio}Comparison of the ratio $\Gamma_{e}/\Gamma_{b}$
of the relaxation rates plotted in Fig. 2(c) of the main text (points)
with the approximation in Eq. (\ref{eq:gebygp}) obtained using dipole
matrix elements (line). }
\end{figure}

Retaining only the piezoelectric term (the dominant term for the parameters
we choose \cite{Hanson2007RMP}) in Eq. (\ref{eq:decayrate}), setting
$\Delta_{d}=\varepsilon_{\mathrm{ph}}=\hbar c_{\nu}k$, and noting
that $I_{\text{ang}}(k)\sim1/k^{2}$ in the limit of large $k$ \cite{Vorojtsov2005},
which is the limit appropriate for the size of $\Delta_{d}$ at small
detuning $ $$|\epsilon|$ {[}see Fig. 2(a) of the main text{]}, we
find $F\left(\Delta_{d}\right)\sim\Delta_{d}^{-1}.$ The ratio of
the decay rates {[}averaged over the nuclear field distribution in
Eq. (7) of the main text{]} can then be estimated as 
\begin{eqnarray}
\frac{\langle\Gamma_{e}\rangle}{\langle\Gamma_{b}\rangle} & \sim & \frac{\langle|d_{e}|^{2}\rangle}{\langle|d_{b}|^{2}\rangle}\left(\frac{\Omega_{z}-\Delta}{\Omega_{z}}\right)^{-1}\nonumber \\
 & = & \left(\frac{\epsilon}{2t}\right)^{2}\left(1+\frac{\Delta}{\Omega_{z}}\right)^{2}\left(1-\frac{\Delta}{\Omega_{z}}\right)^{-1}.\label{eq:gebygp}
\end{eqnarray}
As $\epsilon\rightarrow0$, $\Delta\rightarrow2t\ll\Omega_{z}$ so
that $\langle\Gamma_{e}\rangle/\langle\Gamma_{b}\rangle$ approaches
$\left(\epsilon/2t\right)^{2}.$ The estimated ratio in Eq. (\ref{eq:gebygp})
agrees well with the ratio $\Gamma_{e}/\Gamma_{b}$ obtained from
Fig. 2(c) of the main text for $\left|\epsilon\right|\ll\Omega_{z}$
(Fig. $ $\ref{decayrateratio}).

\bibliographystyle{apsrev4-1}
\bibliography{ECSRDQD}